\documentclass[aps,prx,reprint,superscriptaddress, longbibliography]{revtex4-1}

\usepackage{ucs}
\usepackage[utf8x]{inputenc}

\usepackage[english]{babel}
\usepackage{color,graphicx}
\usepackage{amsmath,amsfonts,amssymb,color,fancybox}
\usepackage{units}

\begin{document}


\title{Signature of a non-harmonic potential as revealed from a consistent shape and fluctuation analysis of an adherent membrane }


\author{Daniel Schmidt}
\affiliation{II. Institut für Theoretische Physik, Universit\"at Stuttgart, 70569 Stuttgart, Germany}

\author{Cornelia Monzel}
\affiliation{Institute of Complex Systems 7: Biomechanics, Forschungszentrum J\"ulich GmbH, 52425 J\"ulich, Germany}
\affiliation{Aix-Marseille Université, CNRS, CINaM UMR 7325, 13288 Marseille, France}
\affiliation{Institute for Physical Chemistry, University of Heidelberg, 69120 Heidelberg, Germany}

\author{Timo Bihr}
\affiliation{II. Institut für Theoretische Physik, Universit\"at Stuttgart, 70569 Stuttgart, Germany}
\affiliation{Institut für Theoretische Physik, Friedrich Alexander Universit\"at Erlangen-N\"urnberg, 91052 Erlangen, Germany}

\author{Rudolf Merkel}
\affiliation{Institute of Complex Systems 7: Biomechanics, Forschungszentrum J\"ulich GmbH, 52425 J\"ulich, Germany}

\author{Udo Seifert}
\affiliation{II. Institut für Theoretische Physik, Universit\"at Stuttgart, 70569 Stuttgart, Germany}

\author{Kheya Sengupta}
\affiliation{Aix-Marseille Université, CNRS, CINaM UMR 7325, 13288 Marseille, France}

\author{Ana-Sun\v{c}ana Smith}
\email{author to whom correspondence should be addressed: smith@physik.uni-erlangen.de}
\affiliation{Institut für Theoretische Physik, Friedrich Alexander Universit\"at Erlangen-N\"urnberg, 91052 Erlangen, Germany}



\date{\today}

\begin{abstract}
The interaction of fluid membranes with a scaffold, which can be a planar surface or a more complex structure, is intrinsic to a number of systems – from artificial supported bilayers and vesicles to cellular membranes. In principle, these interactions can be either discrete and protein mediated, or continuous. In the latter case, they emerge from ubiquitous intrinsic surface interaction potentials as well as nature-designed steric contributions of the fluctuating membrane or from the polymers of the glycocalyx. Despite the fact that these nonspecific potentials are omnipresent, their description has been a major challenge from experimental and theoretical points of view. Here we show that a full understanding of the implications of the continuous interactions can be achieved only by expanding the standard superposition models commonly used to treat these types of systems, beyond the usual harmonic level of description. Supported by this expanded theoretical framework, we present three independent, yet mutually consistent, experimental approaches to measure the interaction potential strength and the membrane tension. Upon explicitly taking into account the nature of shot noise as well as of finite experimental resolution, excellent agreement with the augmented theory is obtained, which finally provides a coherent view of the behavior of the membrane in a vicinity of a scaffold.
\end{abstract}


\maketitle

\section{Introduction}
Phospholipid membranes in cellular and biomimetic systems exhibit significant fluctuations \cite{brochard1975, evans1990, zidovska2006, pelling2007, auth2007a, pierres2008}, which may be of thermal origin, or may arise as a result of active processes in the environment \cite{safran2005,betz2009,hampoelz2011,loubet2012}. Fluctuations play an important role in the regulation of the cell recognition process \cite{pierres2008}, and regulate the adhesiveness of membranes \cite{SS2014}. In the context of protein-mediated interactions, an important role of the fluctuations is to rescale the binding affinity for the macromolecular complexation \cite{schmidt2012} and to promote correlations between the binders, both in the plane of the membrane and while binding to surrounding scaffolds \cite{fenz2011,reister2011}. However, even the qualitative understanding of these processes is a challenge, while the quantitative description is in the nascent stage, and a very active field of research \cite{huppa2010,huang2010}.

The physical framework explaining the thermal membrane fluctuations was provided by Helfrich \cite{helfrich1973} who was the first to calculate the wave-vector dependent fluctuation amplitude as a decreasing function of the membrane stiffness. Shortly after, the effects of the tension originating from the finiteness of the cell or vesicle shape were introduced (for review see \cite{seifert1997} and references therein), even though the precise definition of the tension is still being scrutinized \cite{fournier2008,neder2010,schmid2011}. Meanwhile, a number of methods have been developed to measure the fluctuations of free membranes \cite{evans1986,auth2007a,groves2007,limozin2009}, mostly in red blood cells \cite{brochard1975,fricke1986,boss2012} and phospholipid giant unilamellar vesicles \cite{meleard1992,raedler1995,manneville2001,smith2008}. These early measurements were in good agreement with the theoretical predictions \cite{lipowsky1995}, and were used to determine the tension and the bending stiffness of the membrane. However, very recent data acquired with unprecedented time and space resolution pointed to potential problems \cite{betz2012}. More specifically, data agreed well with the Helfrich model only after the viscosity of the surrounding fluid was set as a parameter, which upon fitting obtained unexpectedly large magnitudes.

Fluctuations of membranes in vicinity of scaffolds, as simple as a hard surface or another membrane, evoked  even more deliberation. In the context of membrane-surface interactions, the focus has often been on specific and discrete protein mediated interactions \cite{weikl2002a,gov2004,auth2005,lin2006a,lin2006b, brannigan2006,krobath2007,krobath2011, hampoelz2011}. However, in addition, there are a number of omnipresent contributions that build a nonspecific potential acting between the two interfaces. Prominent examples of these continuous potentials are the repulsion of the polymers in vesicles, and of the glycocalyx of a living cell. Even more generic are Coulomb and hydration forces \cite{pincus1990}. Equally important contributions to the inter-membrane or membrane-substrate potential are the steric Helfrich repulsion and van der Waals attraction \cite{safinya1986,tanaka2005}, but depending on the system, other potentials may also be involved. The presence of this ubiquitous nonspecific potential of course impacts  the membrane fluctuations \cite{lipowsky1989}, which was well explained close to the unbinding transition \cite{lipowsky1986,netz1995b,manghi2010}. When the system is below the critical temperature, a minimum in the potential is found to appear at finite distances \cite{netz1995a}, from few up to $150$ nanometers interfacial separations \cite{raedler1995,monzel2009}.

The nonspecific membrane-substrate interactions have been studied in adherent vesicles \cite{raedler1995,monzel2009}. The difficulty is, however, that during the spreading of the vesicle in a wetting-like process \cite{sackmann2002}, the tension in the vesicle increases, renormalizing the membrane fluctuations and thus the repulsive contribution to the effective potential \cite{seifert1995b}. In turn, this may affect the position of the minimum of the potential and its strength. Since both are coupled to the vesicle tension, all these parameters must be, in principle, determined self consistently \cite{netz1995a,seifert1995b,mecke2003}, as a function of the membrane stiffness. However, this coupling is still not fully understood when the system is of a finite size and away from the unbinding transition.

The effects of direct membrane-substrate interactions were introduced to theoretical modeling by a harmonic potential, whose strength and position are defined by the curvature and the position of the original  potential, respectively \cite{lipowsky1995,bruinsma1994}. From there on, this harmonic approximation has been used regularly in membrane  adhesion studies \cite{raedler1995,brown2008,speck2011,bihr2012}, even though the range of validity of this approximation has not been experimentally  verified. Furthermore, the above described interplay requires simultaneous determination of the tension and the potential strength. However, after first encouraging attempts \cite{raedler1995, brown2008}, this task has not been fulfilled successfully until now due to limitations of available experimental techniques.

We developed an experimental model system with giant unilamellar vesicles where the membrane is pinned in a controlled geometry, resulting in square shaped segments within which the membrane-substrate interaction is purely nonspecific \cite{monzel2009,monzel2012}. In this geometry, the membrane shape and fluctuations can be measured easily with Dual Wavelength Reflection Interference Contrast Microscopy \cite{monzel2009,monzel2012}, in our setup with an exposure time of $\unit[51]{ms}$, vertical resolution of $\unit[5]{nm}$, and $\unit[100]{nm}$ pixel size. Because the size of the patterned square is much larger than the lateral correlation length of the membrane \cite{raedler1995}, the membrane in the central part of the square is flat on average, and fluctuates around the minimum of the membrane-substrate interaction potential. As such, this system is ideal to explore the nature and consequences of the nonspecific  membrane-substrate interactions, and to test the framework of the available theoretical models. However, for quantitative comparison of theory and experiments, finite time and space resolution of the experimental setup need to be integrated into the theoretical analysis.

In this work, we first provide a general theoretical framework to describe the measured fluctuation amplitudes in adherent  membranes, taking into account the finite space and time resolution of the setup. This allows us to extract the true fluctuations from the measured apparent fluctuations. We then develop a procedure for determining the membrane tension and the strength of the membrane-substrate interaction potential. Three independent approaches are described – by analysis of the shape of the membrane within a grid, by analyzing the spatial correlation function or by analysis of the time correlation function. The three approaches yield very similar results with a very good accuracy, independent of choice of the measurable. We show that for a holistic description it is imperative to go beyond the limitations of the harmonic approximation, which particularly affects the membrane average shape. Consequently, we obtain the first coherent view of the behavior of the membrane in a vicinity of a substrate.

\begin{figure}
 \includegraphics[width=0.95\linewidth]{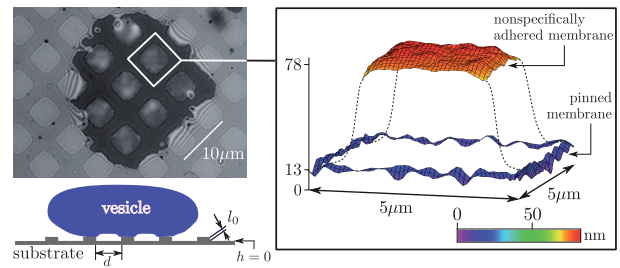}
 \caption{RICM image of a vesicle pinned to a patterned substrate, with squares within which the membrane fluctuates in the nonspecific potential is presented on the left, and schematically below. The reconstruction of the average membrane shape within one square is shown on the right. Only segments of nearly planar membrane were processed to maintain accuracy in the height reconstruction \cite{limozin2009}. The color code indicates the height above the substrate positioned at $h=0$ while $l_0$ denotes the thickness of the adhesive pattern on the glass substrate. The vertical axis in the right figure is in units of nm.}
 \label{fig1}
\end{figure}

\section{Experimental Setup}
\subsection{Materials}
Giant unilamellar vesicles (GUVs) and micropatterned substrates were prepared as described before \cite{monzel2009,monzel2012}. In brief, GUVs composed from SOPC doped with $\unit[2]{mol\%}$ DOPE-PEG2000 and $\unit[5]{mol\%}$ DOPE-cap-biotin (Avanti Polar lipids, USA) were prepared by electro-swelling, and are expected to have a membrane bending stiffness of $\kappa=20 k_\mathrm{B}T$ \cite{evans1990}. Here, $k_\mathrm{B}$ is the Boltzmann constant and $T$ is the temperature. Substrates were prepared by micro-contact printing of BSA-biotin in the form of square grids on ultra-clean class coverslides, yielding an average layer thickness of $\unit[12]{nm}$.  The space within the grid, area $\unit[4]{\mu m} \times \unit[4]{\mu m}$, was back filled with BSA to provide a passive background. The grid itself was further functionalized with neutravidin (NAV). As a result, the biotin in the vesicle membrane binds to the neutravidin on the grid, which then pins the membrane to the pattern, leaving it only subject to the non-specific membrane-substrate potential within the square and typically spreading over several squares, as observed by RICM (Fig. \ref{fig1}).

\subsection{Imaging and Observation}
GUV-substrate interaction was quantified using Dual Wavelength Reflection Interference Contrast Microscopy as described before \cite{limozin2009,monzel2009,monzel2012}. The data was acquired on an inverted microscope (Zeiss Axiovert 200, Carl  Zeiss, G\"ottingen, Germany) equipped with a metal halogenide lamp (X-Cite, Exfo, Quebec, Canada), a dual-wavelength interference filter ($\unit[546]{nm}$ and $\unit[436]{nm}$) and a filter cube with crossed polarizers for illumination; a $63 \times$ Antiflex Plan-Neofluar oil objective; and two separate but synchronized CCD cameras (sensicam qe, PCO, Kehlheim, Germany) for detection in the two wavelength channels. The numerical aperture of illumination was set to $0.54$. Typically, $2000$ consecutive micrographs with a frame rate of $\simeq\unit[20]{Hz}$ were recorded.


\subsection{Analysis}
The recorded intensity images in each frame were converted to height maps following the procedure described previously \cite{monzel2009,monzel2012}. This formalism takes into account all scaffold layers at which refraction occurs, the finite illumination aperture and removes the ambiguities arising from the periodic nature of the intensity to height relationship. Ambiguities arising from camera noise in a given pixel were accounted for by requiring space and time continuity \cite{monzel2009}. The shape of the membrane patches (averaged over $1250$ frames) can be extracted from this analysis (Fig. \ref{fig1}). The height fluctuations for each pixel (defined as the standard deviation of the height from the average, over $1250$ frames) can then be extracted.

\subsection{Spatio-temporal resolution}
The time resolution in this setup is limited by the camera speed and, for the present set of data is $\unit[51]{ms}$. The lateral, in-plane resolution is about $\unit[0.25]{\mu m}$. The pixel size of $\unit[0.1]{\mu m}$ corresponds to slight oversampling which is advantageous for digital image processing, allowing localization precision of single objects of known shape to about $\unit[0.1]{\mu m}$. The vertical resolution is set by the camera noise. The camera noise in this setup is dominated by the statistical shot noise which is proportional to the square root of the intensity \cite{limozin2009,monzel2012}. Typical out of plane resolution is $\unit[5]{nm}$.

\begin{figure}
 \includegraphics[width=0.95\linewidth]{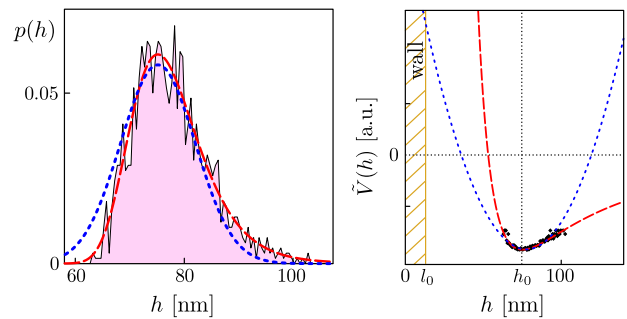}
 \caption{Experimental height probability distribution (black line), and the respective effective potential (black symbols) are shown on the left and right, respectively. Fitting the data with a potential of the Mie or the harmonic form (right), and their Boltzmann factors (left), yield the red dashed and the blue dotted curves, respectively.}
 \label{fig2}
\end{figure}

\section{Theoretical Foundation}
We consider a membrane of bending stiffness $\kappa$ and projected area $S$ put under tension $\sigma$ in the vicinity of a flat substrate. The membrane profile is parameterized in the Monge representation, whereby the membrane height $h({\bf x})$ is determined for every vector ${\bf x}$ residing in the plane of the substrate. Hence, the Hamiltonian of the system can be written in the standard fashion
\begin{equation}
 \mathcal H = \int\limits_S \mathrm d {\bf x} \left ( \frac \kappa2 \left ( \nabla^2 h({\bf x}) \right )^2 + \frac \sigma2 \left ( \nabla h({\bf x}) \right )^2 + V({\bf x}) \right ).
 \label{eq:Hamiltonian}
\end{equation}
The first term in eq. \ref{eq:Hamiltonian} is the contribution due to the bending of the membrane. The second term accounts for the surface tension while the last term in eq. \ref{eq:Hamiltonian} is related to the membrane-substrate interaction potential $V({\bf x})$. Due to the Helfrich repulsion, this potential diverges at short distances and is dominated by attractive van der Waals interactions at large separations. At intermediate distances other contributions to the potential may be significant. Nevertheless, a minimum typically appears at an intermediate height $h_0$, so far reported in the range between $\unit[5]{nm}$ and $\unit[150]{nm}$ above the substrate \cite{limozin2009, tanaka2005,bihr2012,monzel2012, smith2006a,smith2010b}. By definition, and independent of its exact form, the direct potential can be related to the height probability distribution at position ${\bf x}$ (Fig. \ref{fig2}, left)
\begin{equation}
 p(h({\bf x})) \sim \int \mathcal D h^\prime ({\bf x}^ \prime) \mathrm e^{-\mathcal H [h^\prime ({\bf x}^ \prime)]/k_\mathrm{B}T} \delta \left ( h^\prime ({\bf x}) - h ({\bf x}) \right ),
 \label{eq:functionalintegral}
\end{equation}
through a functional integral over all possible membrane profiles weighted by the Boltzmann factor (see SI to \cite{bihr2012}).

The above probability distribution can be measured and used to extract the signature of an effective substrate-membrane potential, the latter being defined as $\tilde V(h)\equiv -k_\mathrm{B}T \ln p(h)$ (right panel of Fig. \ref{fig2}). Within such a construction, the curvature of the minimum of this effective potential $\tilde V^{\prime \prime}(h_0)$ decreases when the fluctuation amplitude $\langle \Delta h^ 2 \rangle$ increases (angle brackets denote ensemble averaging), while it depends on all parameters of the entire Hamiltonian, comprising the direct potential, the tension and the membrane stiffness.

In the current setup, the height probability distribution is obtained by sampling the heights of a small membrane segment in the middle of the square geometry (Fig. \ref{fig1}), to avoid effects of the boundaries. Here, the image was typically averaged over a grid of size $5 \times 5$ pixels, to reduce effects of the camera noise. This height probability distribution has been evaluated in the literature in more complex systems involving ligand-receptor mediated adhesion \cite{smith2006a} or membranes composed of tertiary mixtures \cite{marx2002}. Therein, a Gaussian distribution of a width given by the mean fluctuation amplitude $\langle \Delta h^ 2 \rangle$ was used to describe the data, pointing to the quadratic form of the underlying Hamiltonian, which then implies a harmonic form of the direct membrane-substrate potential. Here, we find for the free membrane segment, small, nevertheless, clear deviations from the Gaussian, whereby fluctuations appear suppressed at the side closer to the substrate (Fig. \ref{fig2}). Since the membrane is nearly flat, the quadratic description used for bending and tension terms seem sufficient, and the only term that can induce deviations from the Gaussian distribution is an anharmonic interaction potential.

A convenient way to account for the anharmonicity of the direct potential is to represent it by the (4,2) Mie-potential
\begin{equation}
 V_\mathrm{M}(h) = \epsilon \left ( \left ( \frac{h_0}{h} \right )^4 - 2 \left ( \frac{h_0}{h} \right )^2 \right ).
 \label{eq:VLJ}
\end{equation}
Here, $\epsilon$ is the strength of the potential in the potential minimum at $h_0$, and the (4,2) structure of $V_\mathrm{M} (h)$ has been chosen to facilitate further numerical calculations. This potential diverges at short distances, and following a minimum, decays algebraically to zero at long distances from the substrate. This captures the key features of the true effective potential. In the two limits (very small and very large distances from the substrate), the Mie potential is, of course, not strictly correct. However, the geometry of the pattern ensures that these two limits are, in practice, not visited by the membrane. Furthermore, the shape of the Mie potential, particularly around the minimum, reproduces the true potential well. Another advantage of the (4,2) potential is that it is defined by only two parameters, which allows a simple comparison with the harmonic potential.

Even though the physics of the problem suggests a more complex potential, theoretical modeling so far has been restricted to the harmonic approximation of the potential (blue dashed curves in Fig. \ref{fig2}) obtained when
\begin{equation}
 V_\mathrm{HA}(h) = \frac \gamma2 \left ( h-h_0 \right )^2
 \label{eq:VHA}
\end{equation}
is used in the Hamiltonian. Thereby, the curvature of the harmonic potential $\gamma$ is the same as that of the Mie form in the minimum yielding
\begin{equation*}
 \gamma = 8\epsilon / h_0^2.
\end{equation*}
The appeal for the harmonic approach does not arise only from the fact that it results in a Hamiltonian with only quadratic terms which is then technically easy to handle, but it maintains consistency between the Monge parameterization (that assumes small curvatures of the membrane) and small (Gaussian) fluctuations around a minimum shape, where each mode is decoupled from others. However, if the membrane is pinned, as it is the case in the patten produced herein, the membrane is significantly moved out of the minimum of the potential. This gives rise to relatively large contributions to the overall energetics of the system, and thus, more accurate treatments of the potential may be required. On the other hand, the nearly flat geometry of the pattern secures the accuracy of the Monge representation, and the fourth order corrections to the bending and tension terms in the Hamiltonian should remain very small. 

In the well established circumstances of the harmonic approximation, the fluctuation amplitude $\langle \Delta h^ 2 \rangle$ is given by \cite{lipowsky1995}
\begin{equation}
 \langle \Delta h^ 2 \rangle = \frac{k_\mathrm{B}T}{(2\pi)^2} \int \mathrm d {\bf q} \frac{1}{\kappa q^4 + \sigma q^2 + \gamma},
 \label{eq:fluct_freemembrane}
\end{equation}
with the notation ${\bf q} \equiv (q_1, q_2)$ and $q\equiv |{\bf q}|$.

\begin{figure}
 \includegraphics[width=0.95\linewidth]{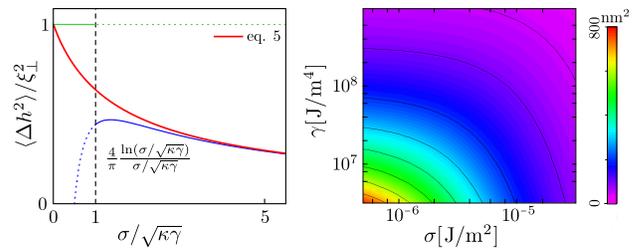}
 \caption{Left: Fluctuation amplitude as a function of the membrane tension (red). Approximations characteristic to the bending and tension dominated regime are shown in green and blue, respectively. Right: Fluctuation amplitude as a function of the tension and potential strength (right). Full lines are contour lines of constant fluctuation amplitude.}
 \label{fig3}
\end{figure}

In the bending dominated regime ($\sigma=0$) the fluctuation amplitude is given by
\begin{equation}
 \xi_\perp^2 \equiv \langle \Delta h^2 \rangle |_{\sigma =0} = \frac{k_\mathrm{B}T}{8\sqrt{\kappa \gamma}},
 \label{eq:xi_perp}
\end{equation}
This particular value for $\xi_\perp^2$ will be referred to as the vertical correlation length henceforth and is shown by the green line in Fig. \ref{fig3} left. In the tension dominated regime, the fluctuation amplitude decays with increasing tension (blue line in Fig. \ref{fig3} left) \cite{lipowsky1995}. However, independently of the parameter range, the fluctuation amplitude depends on both, the potential strength and the tension (Fig. \ref{fig3} right). Consequently, additional information to the fluctuation amplitude is necessary to unambiguously determine $\sigma$ and $\gamma$.

One property that is attainable from the experiment is the equilibrium shape of the membrane itself. We reconstruct the shape from the measured data and compare it to a shape calculated theoretically by minimizing the Hamiltonian, eq. (\ref{eq:Hamiltonian}). In this case, one could expect that the choice of the interaction potential between the membrane and the substrate may have significant influence on the obtained result, simply because the harmonic approximation highly underestimates the repulsion in the proximity of the substrate. On the other hand, small deviations from the Gaussian distribution (Fig. \ref{fig2}) suggest that the fluctuations of the membrane far away from the boundaries could still be treated within the harmonic approximation. These fluctuations may be evaluated through the height probability distribution as shown above, or through the time correlation function at a given position~${\bf x}$
\begin{equation}
 \langle \Delta h({\bf x}, t) \Delta h({\bf x}, 0) \rangle = \frac{k_\mathrm{B}T}{(2\pi)^2} \int \mathrm d {\bf q} \frac{\mathrm e^{-\Gamma(q) t}}{\kappa q^4 + \sigma q^2 + \gamma}.
 \label{eq:time_correlation_free}
\end{equation}
Here, $\Gamma(q)$ are mode dependent damping coefficients for a membrane fluctuating in a potential close to a wall \cite{seifert1994a}
\begin{align}
 \label{eq:dampingcoefficients}
 \Gamma(q) & =  \frac{(\kappa q^4 + \sigma q^2 + \gamma)}{4\eta q} \times \\
 & \times \frac{2 \left ( \sinh (qh_0)^2 - (qh_0)^2 \right ) }{\sinh(qh_0) - (qh_0)^2 + \sinh(qh_0)\cosh(qh_0) + (qh_0)}, \nonumber
\end{align}
with $\eta$ being the viscosity of the surrounding fluid.

For the following, we define the lateral correlation length $\xi_\parallel$ and the characteristic correlation time $\tau^*$ of the membrane fluctuations
\begin{equation}
 \xi_\parallel \equiv \sqrt[4]{\kappa/\gamma},\text{    } \tau^* \equiv \eta/\sqrt[4]{\kappa \gamma^3}.
 \label{eq:xi_parallel}
\end{equation} 
These values for a tensionless membrane provide the lower ($\xi_\parallel$) and upper ($\tau^*$) bound for the lateral correlation length and the correlation time, respectively, in the presence of tension. For typical experimental settings they amount to $\xi_\parallel\simeq \unit[200]{nm}$ and $\tau^* \simeq \unit[0.1]{ms}$.
Under these circumstances, the correlation function given in eq. \ref{eq:time_correlation_free} can be accurately evaluated only in the central segment of the free membrane patch. Because of the potential influence of the boundaries in the square geometry, two-point spatial correlations are not discussed.

\begin{figure}
 \includegraphics[width=0.95\linewidth]{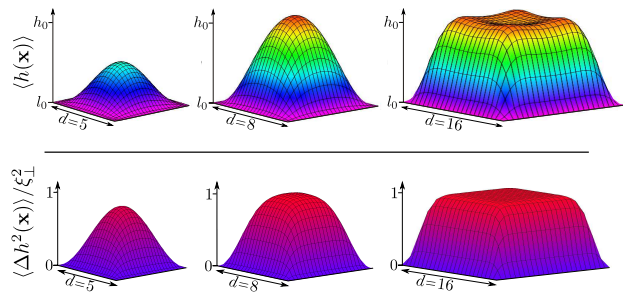}
 \caption{Membrane in a nonspecific harmonic potential with the minimum at the height $h_0$, pinned to a square of an edge length $d$ (in units of $\xi_\parallel$). The universal mean membrane shape and the associated profile of the mean fluctuation amplitude (normalized by $\xi_\perp^2$), are shown in the top and bottom rows, respectively. All profiles are calculated for $\kappa=\unit[20]{k_\mathrm{B}T}$, $\sigma=0$, $\gamma= \unit[2 \cdot 10^7]{J/m^4}$.}
 \label{fig4}
\end{figure}

\section{Methods}
In this section we focus on the development of methods which allow the comparison of theoretical models and experimental measurables \cite{monzel2012}. In particular, we calculate the shape of the membrane and relate the true correlation functions to apparent ones, which differ due to finite resolution of the experimental setup. However, the final, experimentally recorded height integrates effects of thermal noise inherent to the data acquisition techniques, which we also account for in our discussions.

\begin{figure*}
 \includegraphics[width=0.95\linewidth]{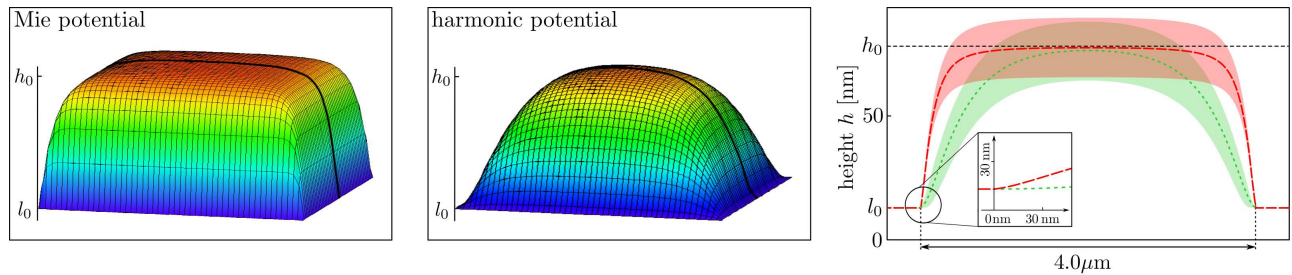}
 \caption{Comparison of the mean membrane profile of a membrane residing in a Mie-potential (left panel) and a harmonic potential of identical curvature (middle panel). The cross-section through the center of the shapes is also shown, enveloped by the mean fluctuations of the shape (right panel). The profile associated with the Mie-potential is shown in red, while the profile in the harmonic potential is shown in green. The inset shows a detailed view of the membrane shape near the pinning point. The shapes are determined for $\sigma= \unit[6.6\cdot 10^{-6}]{J/m^2}$ and $\gamma=\unit[3.3\cdot 10^7]{J/m^4}$, and $h_0-l_0=\unit[65]{nm}$ as for vesicle segments shown later in Fig. \ref{fig10}.}
 \label{fig5}
\end{figure*}

\subsection{Calculation of the membrane shape}
The equilibrium shape $\langle h({\bf x})\rangle$ of the membrane has to fulfill the boundary conditions
\begin{equation}
 \langle h({\bf x})\rangle|_{\partial S} = l_0 \qquad \text{and} \qquad \nabla \langle h({\bf x})\rangle|_{\partial S}=0.
 \label{eq:boundary_conditions}
\end{equation}
The first condition fixes the height of the membrane at the edge of the square frame of a surface $S$. The second condition ensures a finite bending energy of the membrane by requiring a zero contact angle along the frame.

For the calculation of the equilibrium shape $\langle h({\bf x})\rangle$ in the harmonic potential (Fig. \ref{fig4}), with the above set of boundary conditions, the equilibrium shape $\langle h({\bf x})\rangle$ is expanded into a set of orthonormalized functions $\Psi_{ij} ({\bf x})$
\begin{equation}
 \langle h({\bf x})\rangle = \sum_{ij} a_{ij}\Psi_{ij}({\bf x}) + l_0,
 \label{eq:ansatz_mean_shape}
\end{equation}
where each $\Psi_{ij} ({\bf x})$ is given by a product of two one dimensional functions, $\Psi_{ij} ({\bf x}) = \psi_i(x_1)\psi_j(x_2)$, with $x_1$ and $x_2$ being components of the position vector ${\bf x}$. Each $\psi_i$ is a stationary solution of the one dimensional Hamiltonian \cite{weikl2003}, and satisfies the relevant boundary conditions. Thus, the membrane shape $\langle h({\bf x}) \rangle$ fulfills the boundary conditions for every possible set of expansion coefficients $\left \{ a_{ij} \right \}$.  The optimum shape is found by minimizing the entire Hamiltonian (eq. \ref{eq:Hamiltonian}) with respect to the entire set $\left \{ a_{ij} \right \}$.

In the Mie-potential, the equilibrium shape cannot be minimized analytically. Therefore, the equilibrium shape is found numerically by discretizing the membrane on a mesh of $100 \times 100$ lattice segments and applying a steepest descent optimization to the membrane shape $\langle h({\bf x})\rangle$

As can be seen from Fig. \ref{fig5}, because the harmonic approximation significantly underestimates the repulsion between the membrane and the substrate, the shape of the profile is significantly different in the two approaches. We find that the harmonic approximation correctly predicts trends in the dependence of the shape on the tension and the potential strength of the membrane, but cannot be used for quantitative understanding of experimentally obtained profiles. Consequently, anharmonic contributions are absolutely necessary to understand the observed fast decay of shapes close to the edge of the pattern.

\subsection{Membrane fluctuations}
The fluctuations in the harmonic potential are calculated in a similar way as the shape: The fluctuations $\Delta h({\bf x},t)$ of the membrane emerge from the instantaneous membrane conformations as small deviations from the equilibrium shape
\begin{equation}
 h({\bf x}, t)= \langle h ({\bf x}) \rangle + \Delta h({\bf x},t).
 \label{eq:fragmentation}
\end{equation}
In order to calculate $\Delta h({\bf x},t)$, the fluctuating profile is expanded into the same set of orthogonal functions as the mean profile
\begin{equation}
 \Delta h({\bf x},t) = \sum_{ij} b_{ij}\Psi_{ij}({\bf x}).
 \label{eq:ansatz_fluctuations}
\end{equation}
The second variation to the Hamiltonian is then related to the total energy of the fluctuations
\begin{equation}
 \delta^2 \mathcal H = \frac 12 \sum_{ijkl} b_{ij} E_{ijkl}b_{kl}
 \label{eq:2nd_variation}
\end{equation}
with $E_{ijkl}$ being the energy arising from coupling the $(ij)$ with the $(kl)$ mode. The mean square deviations from the average shape fulfil the equipartition theorem, $\langle b_{ij}b_{kl}\rangle = k_\text{B}T (E^{-1})_{ijkl}$, and thus
\begin{equation}
\langle \Delta^2 h({\bf x}) \rangle = k_\mathrm{B}T \sum_{ijkl} \Psi_{ij}({\bf x}) (E^{-1})_{ijkl} \Psi_{kl}({\bf x}).
\label{eq:ansatz_fluctampl}
\end{equation}
Consequently the profile of the mean squared fluctuation amplitude can be evaluated numerically by determining the tensor $E_{ijkl}$ (Fig. \ref{fig4}).

For determining the membrane fluctuations in the Mie-potential we use the same approach as for the fluctuations in the harmonic potential, which requires finding the second variation of the Hamiltonian with respect to the appropriate equilibrium shape (e.g. left panel in Fig. \ref{fig5}). In the current case, $V_{M}(h)$ is no longer harmonic and the curvature of the potential affecting membrane fluctuations depends on the height that the membrane achieves along the profile. To obtain the second variation of a given profile $h({\bf x})$, we thus expand $V_{M}(h)$ in orders of $\Delta h({\bf x},t)$, which results in $\delta^2 \mathcal H$ that is of the form as in eq. \ref{eq:2nd_variation} which contains implicitly a distance dependent $\gamma_{M}$ given by the scaling function $g \left ( \langle h({\bf x})\rangle \right )$
\begin{align}
 \gamma_{M} & \equiv V^{\prime \prime}_{M} \left ( \langle h({\bf x}) \rangle \right ) = \gamma g\left (\langle h({\bf x})\rangle \right ) \nonumber\\
 & = \frac{\gamma h_0^ 2}{4 \langle h({\bf x})\rangle^ 2} \left [ 10 \left ( \frac{h_0}{\langle h({\bf x} \rangle} \right )^4 - 6 \left ( \frac{h_0}{\langle h({\bf x} \rangle} \right )^2 \right ].
 \label{eq:gamma_LJ}
\end{align}
For the membrane resting in the minimum $h_0$ the scaling function becomes unity and thus, the fluctuations of a unbound membrane in the $V_{M} (h)$ are exactly the same as in the harmonic potential. For any height of the membrane somewhere between $h_0$ and $l_0$ the scaling function significantly increases the effective potential strength $\gamma_{M}$, resulting in strongly suppressed fluctuation amplitudes when the membrane deviates from the minimum of the potential (right panel in Fig. \ref{fig5}).

\subsection{Accounting for the finite resolution of the acquisition system}
\begin{figure}
 \includegraphics[width=0.95\linewidth]{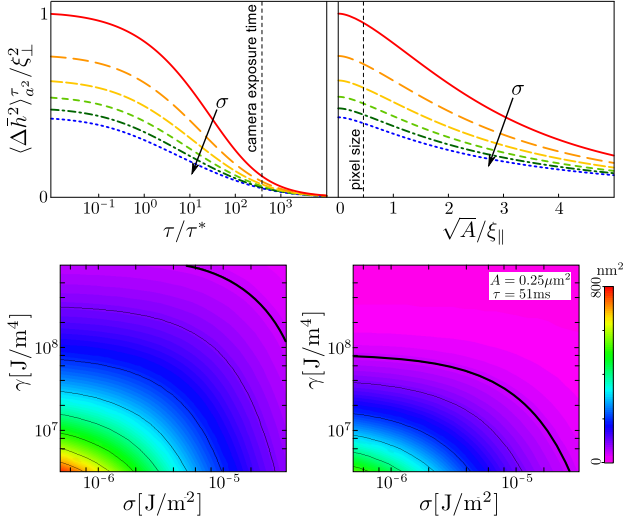}
 \caption{Normalized apparent fluctuation amplitude as a function of temporal (upper left panel) and spatial resolution (upper right panel). The tension increases (in the direction of the arrow) from $\sigma=0$ to $\unit[5]{\sqrt{\kappa \gamma}}$. The true and the apparent fluctuation amplitudes are shown in bottom left and bottom right panels, respectively. The latter has been calculated for a fixed integration time of $\tau=\unit[51]{ms}$ and is averaged over an area of $A=\unit[0.5]{\mu m}\times \unit[0.5]{\mu m}=\unit[0.25]{\mu m^2}$. Contour lines of constant mean square fluctuation amplitudes are indicated. Specifically, the contour lines of the true and apparent fluctuation amplitude of $\unit[50]{nm^2}$ are presented by the thick lines.}
 \label{fig6}
\end{figure}

Measuring membrane fluctuations is the key to determining the physical parameters of the system \cite{brochard1975,mecke2003,bruinsma1994}. However, due to the finite temporal and spatial resolutions of the experimental techniques, only apparent fluctuation amplitudes are measured which may significantly differ from true fluctuations of the membrane (Fig. \ref{fig6}).

Due to a finite time resolution, modes with a life time smaller than $\tau^*$ cannot be detected. In the current system, spatial resolution almost matches the lateral correlation length $\xi_\parallel$ whereas the integration time by far exceeds the correlation time $\tau^*$. Therefore, temporal integration has a particular large effect (Fig. \ref{fig6}). Certain specialized techniques can make faster recording of intensity fluctuations \cite{betz2009,pecreaux2004,franosch2011}, but these acquire the information on the state of the membrane only in a single point and are not compatible with spatial imaging of the membrane. Consequently, developing procedures to interpret the measured fluctuations become imperative.

The effects of the temporal resolutions were first taken into account for the spectra obtained from measuring the fluctuations of the contour of a freely suspended giant vesicle \cite{faucon1989}. Thereby, the vesicle shape was parameterized by spherical functions and the temporal average of the time-dependent correlation function was performed. Here we adapt this procedure to a situation where a flat segment of the membrane parameterized in Monge representation fluctuates close to the wall. Consequently, the effects of the nonspecific potential are taken into account, and averaging is performed with the appropriate damping coefficients, given by eq. \ref{eq:dampingcoefficients}.

The spatial averaging occurs due to a finite lateral resolution of the experiment. The camera averages the signal over an area $A$ and only fluctuation modes with a wave-length larger than $\sqrt{A}$ can be fully resolved.

As discussed, smearing the true membrane height at the position ${\bf x}$ and at the time $t$ gives rise to the apparent membrane height $\bar h _A^\tau ({\bf x},t) $, whereby the subscript $A$, and the superscript $\tau$ indicate the spatial and temporal integration of the given measurement, expressed in square microns and milliseconds, respectively
\begin{equation}
 \bar h _A^\tau ({\bf x},t) = \int\limits_0^\tau \frac{\mathrm d t^\prime}{\tau} \int\limits_A \frac{\mathrm d {\bf x}^\prime}{A} h({\bf x} + {\bf x}^\prime, t+t^\prime).
 \label{eq:h_finres}
\end{equation}
From eq.  \ref{eq:h_finres} it is straightforward to derive the apparent time correlation function
\begin{align}
 \langle \Delta \bar h ({\bf x},t) \Delta \bar h({\bf x}, 0) \rangle _A^ \tau = \int\limits_0^ \tau \int\limits_0^ \tau \frac{\mathrm d t_1^ \prime \mathrm d t_2^\prime}{\tau^ 2} \iint\limits_A \frac{\mathrm d{\bf x}_1^ \prime \mathrm d{\bf x}_2^ \prime}{A^ 2} \nonumber \\
 \langle \Delta h ({\bf x}+{\bf x}_1^ \prime,t+t_1^\prime) \Delta h({\bf x}+{\bf x}_2^ \prime, t+t_2^ \prime) \rangle.
 \label{eq:time_correlation_finres}
\end{align}
 The true time correlation function $\langle \Delta h ({\bf x},t) \Delta h({\bf x}, 0) \rangle$ in real space is given in eq. \ref{eq:time_correlation_free}. The apparent time correlation function is found in Fourier space
\begin{align}
 & \langle \Delta \bar h ({\bf x},t) \Delta \bar h({\bf x}, 0) \rangle _A^ \tau\nonumber\\
 & \quad = \frac{k_\mathrm{B}T}{(2\pi)^2} \int \mathrm d {\bf q} \frac{\mathrm e^{-\Gamma(q) t}}{\kappa q^4 + \sigma q^2 + \gamma} \phi_A({\bf q}) \psi^ \tau({\bf q}),
 \label{eq:time_correlation_finres2}
\end{align}
as the convolution of the true correlations with the effects of the temporal and spatial averaging. Here, $\psi^ \tau({\bf q})$ is a function of the time component
\begin{align}
 \psi^ \tau({\bf q}) & \equiv \int\limits_0^ \tau \int\limits_0^ \tau  \frac{\mathrm d t_1^\prime \mathrm d t_2^\prime}{\tau^2} \mathrm e^{-\Gamma ({\bf q}) |t_1^\prime - t_2^\prime|} \nonumber \\
 & = \frac{\mathrm e^{-\Gamma(q)\tau} - 1 + \Gamma(q) \tau}{\Gamma^2(q)\tau^2}
 \label{eq:time_averaging}
\end{align}
and $\phi_A({\bf q})$ of the spatial component
\begin{align}
 \phi_A({\bf q}) \equiv \iint \limits_A \frac{\mathrm d {\bf x}_1^\prime \mathrm d {\bf x}_2^\prime}{A^2} \mathrm e^{-i{\bf q} ({\bf x}_1^\prime - {\bf x}_2^\prime)}.
 \label{eq:spatial_averaging}
\end{align}
In principle, one could use any form of the patch $A$. In the special case of a square geometry of the adhesion pattern, it is convenient to follow the boundaries and keep the square geometry for the averaging procedure, which results in
\begin{equation}
 \phi_A({\bf q}) = \frac{16}{A^2}\frac{\sin \left ( \frac{\sqrt{A}q_1}{2} \right )^2 \sin \left ( \frac{\sqrt{A}q_2}{2} \right )^2}{(q_1q_2)^2}.
 \label{eq:spatial_averaging_square}
\end{equation}
For perfect temporal resolution $\tau \to 0$ and $\psi^0({\bf q}) \to 1$. Likewise, for perfect spatial resolution $A\to 0$ and $\phi_0({\bf q})\to 1$. For objects of known shape one could improve this procedure by using a more complex optical resolution function to deconvolute correlations between neighboring pixels \cite{wiegand1998}.

The apparent mean square fluctuation amplitude easily emerges from eq. \ref{eq:time_correlation_finres2} for $t=0$ as
\begin{equation}
 \langle \Delta \bar h^2 \rangle _A^ \tau  = \frac{k_\mathrm{B}T}{(2\pi)^2} \int \mathrm d {\bf q} \frac{1}{\kappa q^4 + \sigma q^2 + \gamma} \phi_A({\bf q}) \psi^ \tau({\bf q}).
 \label{eq:fluctampl_finres}
\end{equation}
In Fig. \ref{fig6}, we show the influence of temporal and spatial averaging of true fluctuations. The results show that the finite resolution of the experiment affects the apparent fluctuation amplitude by making it systematically smaller. Furthermore, we find that the limitations of the experimental technique have larger effects in systems subject to larger tensions and stronger interaction potentials.

\begin{figure}
 \includegraphics[width=0.95\linewidth]{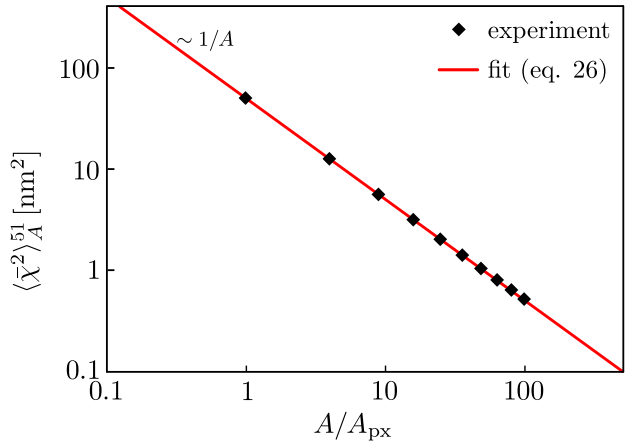}
 \caption{The mean square amplitude of the camera noise averaged over an area $A$ containing $N$ pixels of area $A_\mathrm{px}$. The noise decreases exceptionally well with $1/A\sim 1/N$, proving the camera noise being independent for different pixels.}
 \label{fig7}
\end{figure}

\subsection{Accounting for the background noise of the acquisition system}
Another effect that impacts the experimental data is that of the background noise $\chi({\bf x}, t)$ of the acquisition system. In case of optical microscopy, the latter arises mostly from the intensity dependent fluctuations in the number of photons reaching the detector. In principle, this shot noise is Poisson distributed, but due to the high number of photons detected in a typical RICM experiments, it can be treated as Gaussian distribution leading to the noise increase with the square root of the intensity \cite{monzel2009, monzel2012}.

As the membrane height and the noise are assumed to be independent, the measured instantaneous membrane profile $\tilde h_A^\tau ({\bf x},t)$ is given by
\begin{equation}
 \tilde h_A^\tau ({\bf x},t) = \bar h_A^\tau ({\bf x},t) + \bar \chi_A^\tau ({\bf x},t).
 \label{eq:fragmentation_hnoise}
\end{equation}
The first term on the right hand side is the apparent height and the second term is the contribution from the apparent noise, whereby the latter emerges from temporal and spatial averaging of the background noise $\chi({\bf x}, t)$, over the time $\tau$ and area $A$, respectively. Similarly, the measured fluctuation amplitude of the membrane $\langle \Delta \tilde h^2 ({\bf x}) \rangle _A^\tau$ is the sum of the apparent membrane fluctuation amplitude, eq. \ref{eq:fluctampl_finres}, and the variance $\langle \bar \chi^2 ({\bf x}) \rangle _A^\tau$ is the ensemble average of the apparent noise
\begin{equation}
 \langle \Delta \tilde h^2 ({\bf x}) \rangle _A^\tau = \langle \Delta \bar h^2 ({\bf x}) \rangle _A^\tau + \langle \bar \chi^2 ({\bf x}) \rangle _A^\tau.
 \label{eq:fragmentation_dh2_noise}
\end{equation}
The time component of the apparent noise goes with $\sqrt{\tau}$, since it scales with the square root of the number of photons detected \cite{monzel2009, monzel2012}.

For a pixilated image, the height can be measured only at discrete positions ${\bf x}_i$, and the spatial resolution imposes the minimum area for averaging to be $A_\mathrm{px}=a^2$. As such, the measured height $\tilde h_{A_\mathrm{px}}^\tau ({\bf x}_i, t)$ of a single pixel inherently incorporates temporal and spatial averaging of noise on a level of a pixel $\bar \chi_{A_\mathrm{px}}^\tau ({\bf x}_i)$, the latter being of a particular background intensity.

We consider a segment of a pixilated membrane of an area $A$. This area can be of arbitrary shape as long as it consists of $N$ pixels of identical background intensity (e.g. identical average height) for which noise is uncorrelated. In that case, the apparent noise is
\begin{align}
 & \langle \bar \chi ({\bf x}_i)^2 \rangle _A^\tau = \frac{1}{N^2} \sum_{{\bf x}_k, {\bf x}_l \in A} \langle \bar \chi _{A_\mathrm{px}}^\tau({\bf x}_i + {\bf x}_k) \bar \chi _{A_\mathrm{px}}^\tau({\bf x}_i + {\bf x}_l) \rangle \nonumber\\
 & = \frac{1}{N^2} \sum_{{\bf x}_k, {\bf x}_l \in A} \langle \bar \chi^2 \rangle_{A_\mathrm{px}}^\tau \delta_{kl} = \frac 1N \langle \bar \chi^2 \rangle_{A_\mathrm{px}}^\tau = \frac{a^2}{A} \langle \bar \chi^2 \rangle_{A_\mathrm{px}}^\tau.
 \label{eq:noise_amplitude}
\end{align}
Here, we sum over all pixels within the considered area $A$. This result shows that averaging over several pixels may decrease the effect of the camera noise to negligible levels. For example, for the current experimental conditions, the variance of the apparent mean square amplitude of the camera noise drops below $\unit[3]{nm^2}$ upon averaging over $25$ pixels (Fig. \ref{fig7}).

\section{Results}
In the following we develop three approaches to simultaneously determine the membrane tension $\sigma$ and the strength $\gamma$ of the membrane-substrate interaction potential. The common denominator to all of the approaches is determining the true mean fluctuation amplitude from the measured one. Thereby, it is assumed that the membrane resides in the minimum of the potential, which is well justified by the flatness of the membrane profile in the shape reconstruction (Fig. \ref{fig1}).

Determining the true fluctuation amplitude isolates the correct contour line (lower panel in Fig. \ref{fig6}). However, additional information is necessary to resolve the interdependence of $\langle \Delta h^2 \rangle$ on $\sigma$ and $\gamma$. Such information can be provided by determining the shape of the membrane or the correlations. In the general case of adherent membranes not all of these parameters can be determined, and the particular availability depends on the particular experimental situation. In the current setup, all possible measures are obtained simultaneously, due to a particular design of the system. This allows us to take one measure at a time, deconvolve $\sigma$ and $\gamma$, and compare the obtained results from different choices. However, if the theoretical model is complete, $\sigma$ and $\gamma$ emerge as independent of the approach. Inability to obtain systematic values of the tension and the potential strength should point either to deficiencies of the theoretical description, or to problems with the experimental technique.

\begin{figure}
 \includegraphics[width=1\linewidth]{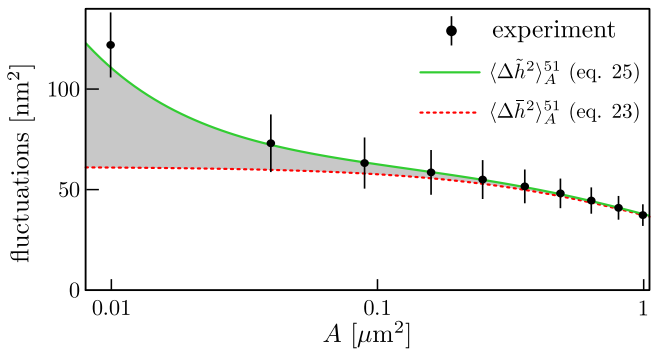}
 \caption{Determining the tension and the potential strength by systematic spatial averaging over a square of area A. The best fit results in $\sigma = \unit[5.0 \cdot 10^{-6}]{J/m^2}$ and $\gamma = \unit[3.7 \cdot 10^{7}]{J/m^4}$.}
 \label{fig8}
\end{figure}

\subsection{Approach 1: Systematic spatial averaging}
Within this approach, the measured fluctuation amplitude $\langle \Delta \tilde h^2 ({\bf x}) \rangle _A^\tau$ of a flat segment of the membrane is determined as a function of the averaging area $A$. Thereby, $A$ is varied by systematically increasing the number of pixels in the observed membrane segment ($1 \times 1$ px, $2 \times 2$ px, $3 \times 3$ px, etc.), around the central pixel in the frame. This results in a square of a length $a=\unit[0.1]{\mu m} \sqrt{N}$, $N$ being the number of pixels, for which the spatially averaged height $\tilde h$ is determined in each instance of time. This provides a sequence from which the mean height $\langle \tilde h \rangle _A^\tau$ and the mean square deviation $\langle \Delta \tilde h^2 ({\bf x}) \rangle _A^\tau$ is determined for each choice of $a$. The obtained data are shown with symbols in Fig. \ref{fig8}. To avoid influences from the boundaries, we restrict the total area of interest to a square of $\unit[1]{\mu m^ 2}$ in the center of the pattern.

To determine the tension and the potential strength, eq. \ref{eq:fragmentation_dh2_noise} is fitted to the data, with $\sigma$ and $\gamma$ being the fit parameters (Fig. \ref{fig8}). Thereby, the contribution from the apparent fluctuation in eq. \ref{eq:fragmentation_dh2_noise} is given by eqs. \ref{eq:time_averaging}, \ref{eq:spatial_averaging_square} and \ref{eq:fluctampl_finres}, whereas the contribution of the noise $\langle \bar \chi^2 \rangle _A^\tau$ was determined independently, for a pixel of the equivalent brightness. If the contribution from the noise were not known a priori, the procedure could be applied with a fit with three free parameters.

For the particular vesicle adhered to a pattern, as shown in Fig. \ref{fig1}, the camera noise was found to be $\langle \bar \chi^2 \rangle _A^\tau = \unit[49]{nm^2}$. The systematic spatial averaging gives $\sigma = \unit[5.0 \cdot 10^{-6}]{J/m^2}$ and $\gamma = \unit[3.7 \cdot 10^{7}]{J/m^4}$. Thereby, the accuracy of the fit provides the mean square amplitudes within the error bar of the experiment.

\begin{figure}
 \includegraphics[width=0.95\linewidth]{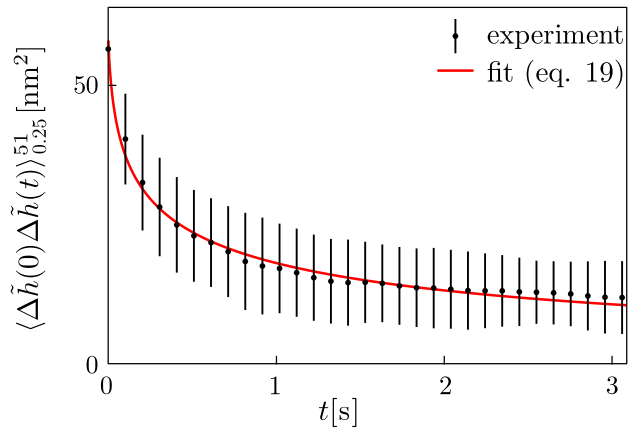}
 \caption{Measured temporal correlation function (every second data point presented) is fitted with the expression given in eq. \ref{eq:time_correlation_finres2}. The fitting procedure provides $\sigma=\unit[13.0\cdot 10^{-6}]{J/m^2}$ and $\gamma=\unit[1.0\cdot 10^7]{J/m^4}$.}
 \label{fig9}
\end{figure}

\subsection{Approach 2: Time dependent correlation function}
Despite the somewhat limited time resolution of the setup, a quite sensitive approach to determining the tension and the nonspecific potential is fitting the measured time dependent  correlation function $\langle \Delta \tilde h ({\bf x},t) \Delta \tilde h({\bf x}, 0) \rangle$. The latter is still sensitive to the spatial resolution. It is instructive to use relatively large segments of the membrane to decrease the effects  of the camera noise. Hence, we typically consider an area consisting of $5 \times 5$ pixels for which the spatially average height is calculated in each instance of time. This provides a sequence  of heights $\tilde h ({\bf t})$ from which the time correlation function is calculated.

The reason for this sensitivity is the course of the time dependent correlation function over a temporal regime (from  $t=0$ to $t<\unit[3]{s}$). Within this range the correlations decay from the fluctuation amplitude  $\langle \Delta \tilde h^2 \rangle$ and ultimately reach zero. Therefore, the fitting curve has to match three characteristics,  the fluctuation amplitude for $t=0$, which is the mean  fluctuation amplitude $\langle \Delta \tilde h^2 \rangle_A^\tau$ , the long-time  behaviour (visible beyond $t\simeq \unit[1]{s}$) and the characteristic decay time (see Fig. \ref{fig9}). These stringent restrictions make it rather simple to find appropriate  parameters.

Data fitting is performed by applying eq. \ref{eq:time_correlation_finres2} with $\sigma$ and $\gamma$ being the free parameters. The best fitting values for averaging  over $N=25$ pixels are $\sigma=\unit[13.0\cdot 10^{-6}]{J/m^2}$ and $\gamma=\unit[1.0\cdot 10^7]{J/m^4}$, for the case of the vesicle discussed in the approach 1.

\subsection{Approach 3: The membrane shape}
The last available free parameter is the very shape of the membrane. The membrane is expected to be flat and in the  minimum of the potential. Hence, we obtain a large segment of a membrane. However, the regions along the pattern (dark areas in Fig. \ref{fig1}) can be regarded as membrane residing in a different state, the  latter being characterised by an effectively much stronger potential with a minimum very close to the substrate. The experimental design imposes the geometry and hence the occurrence of the two states. The transition of the membrane between the two states occurs within the pattern, providing a membrane interface that is in principle subject only to nonspecific interactions. Because the height difference between the two states is of the order of $\unit[50]{nm}$, the deviations from the minimum of the effective nonspecific potential can no longer be regarded as small, necessitating the systematic use of the Mie-potential.

\begin{figure}
 \includegraphics[width=0.95\linewidth]{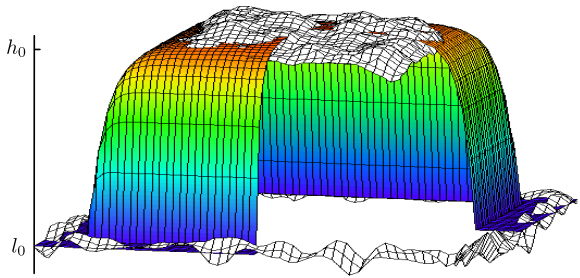}
 \caption{Fitting the theoretically obtained mean shape (colored) to the experimentally determined mean profile of the membrane (gray). The best fit is found for with $\sigma=\unit[6.6\cdot 10^{-6}]{J/m^2}$ and $\gamma=\unit[3.0\cdot 10^7]{J/m^4}$.}
 \label{fig10}
\end{figure}

The fitting procedure is performed in two steps. We first determine the mean fluctuation amplitude of the membrane $\langle \Delta \tilde h^2 \rangle _A^\tau$ in the center of the weakly adhered fragment of the membrane. Thereby, it is preferable to choose large $A$ ($A=\unit[0.25]{\mu m^2}$) to avoid effects of the camera noise. Determining the mean fluctuation amplitude reduces the choice of $\sigma$ and $\gamma$ to a particular subset of values presented by the relevant contour line (Fig. \ref{fig6}). In the second step, a family of shapes with $\sigma$ and $\gamma$ along the contour line are calculated and the shape of the smallest mean square deviation from the experimental shape is determined. The best fitting shape ascertains $\sigma$ and $\gamma$, whereby no additional constraints were imposed. For the vesicle treated in the approach 1 and 2, this procedure provides the shape shown in Fig. \ref{fig10}, associated with $\sigma=\unit[6.6\cdot 10^{-6}]{J/m^2}$ and $\gamma=\unit[3.0\cdot 10^7]{J/m^4}$.

From the experimental point of view, it is only possible to reconstruct shapes of sufficient planarity at this stage. However, this affects the model reconstruction only slightly. Because the membrane in both adhesion states reaches the minimum of the potential at zero angles, the large section of the steep profile must be nearly linear. Hence, obtaining the width of the interface is almost equivalent to determining the overall shape. Here, the strong repulsion from the substrate in the Mie-potential promotes steep interfaces, which is not the case for the harmonic potential. It is also worth noticing that the camera noise has no effect on the measured mean shape since $\langle \chi \rangle =0$.

\begin{figure}
 \includegraphics[width=0.95\linewidth]{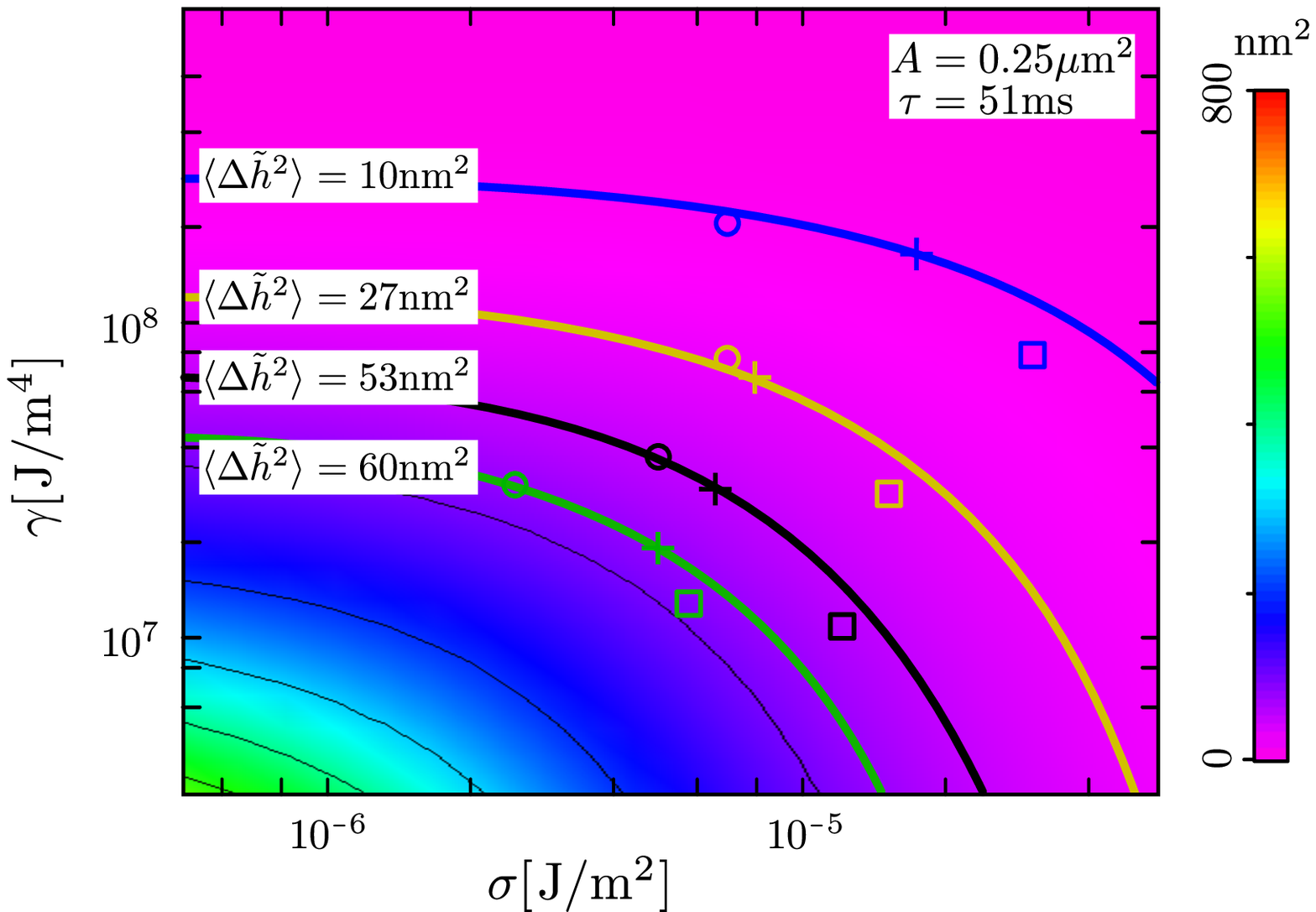}
 \begin{tabular}{|l|c|c|}\hline
  {\bf Method} & $\  \sigma$ [$10^{-6}$J/m$^2$]$\ $ & $\ \gamma$ [$10^{7}$J/m$^4$]$\ $ \\ \hline
  Systematic spatial averaging $\ $ & $5.0$ & $3.7$ \\ \hline
  Time correlations & $13.0$ & $1.0$ \\ \hline
  Shape fitting & $6.6$ & $3.0$ \\ \hline
 \end{tabular}
 \caption{Comparison of fitting procedures for a single vesicle is shown in the table and for four vesicles in the graph. The vesicle discussed in the manuscript and fittings shown in Figs. \ref{fig7} - \ref{fig10} is indicated by black lines and symbols. Circles denote results of determining the potential strength and the tension by the method of systematic averaging, while crosses and squares are obtained by fitting the shape and the time dependent correlation function, respectively. Results of all fitting procedures for each vesicle lie very close to the appropriate contour line.}
 \label{fig11}
\end{figure}

\section{Discussion}
In this work, we presented three independent methods to determine the strength of the nonspecific potential and the tension of membranes that weakly adhere in homogeneous potentials. All three methods were applied to the same sets of data allowing for the first time, to our knowledge, the direct comparison between various approaches. After accounting for experimental limitations, all procedures provide values within the same order of magnitude for both the tension and the interaction potential strength, as can be seen in Fig. \ref{fig11} and the related table. This is particularly important for the determination of the relevant parameters in more complex experimental situations where only one of these procedures can be used depending on the circumstances.

\begin{figure*}
 \includegraphics[width=0.95\linewidth]{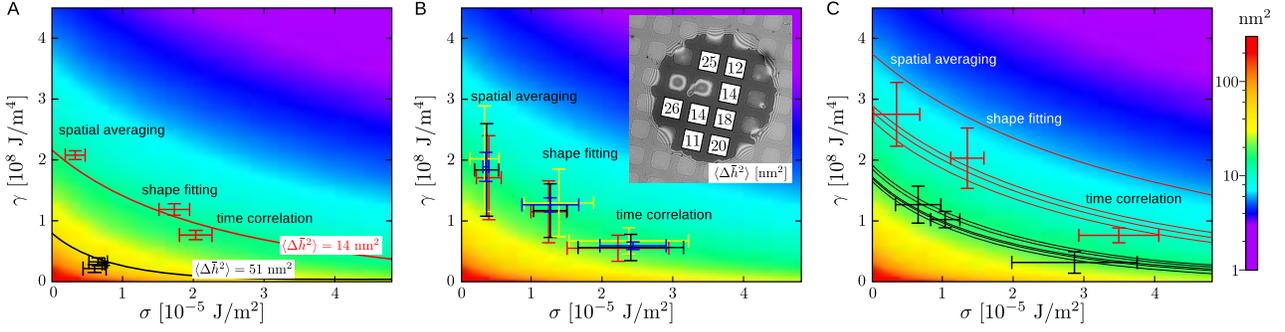}
 \caption{Analysis of reliability of the three presented approaches for extracting the potential strength and the tension in the membrane. The apparent fluctuation amplitude as a function of the potential strength and tension is shown as a background of each graph. The color code is given on the right. Contours associated with the apparent mean fluctuation amplitudes that are particular to each data set are displayed with full lines. In all panels, the spatial and temporal averaging is performed with $A=\unit[0.25]{\mu m^2}$ and $\tau =\unit[0.51]{ms}$, respectively. (A) Mean values and standard deviations (error bars) associated with performing the analysis on several data recordings from the same patch. Results are presented for one patch with the low (red) and one patch with a high (black) mean apparent fluctuation amplitude (indicated by numbers on the relevant contour line). (B) Mean values and standard deviations associated with averaging the fit results over the eight patches on one vesicles The patches and the vesicle is shown in the inset by white square. The numbers within the squares indicate the apparent mean membrane fluctuation amplitude $\langle \Delta \bar h^2 \rangle$ for the particular membrane segment in units of nm$^2$. The analysis was performed on two short data sequences (two sub-samples shown in red and yellow) and on a long sequence (black symbols). The results of averaging over three squares with the mean apparent fluctuation amplitude between 14 and 18 $nm^2$ is shown in blue. (C) Mean values and standard deviations associated with averaging over fitting results from membrane patches with similar mean apparent fluctuation amplitudes on different vesicles. The results for patches with small fluctuation amplitudes, ranging from 6 to 11 $nm^2$, are shown in red while the results for large fluctuations, ranging from 18 to $\unit[20]{nm^2}$, are shown in black.}
 \label{fig12}
\end{figure*}

To estimate the reliability of each approach, we first split the data into several sub-samples, i.~e. shorter time sequences of the membrane height, and perform the analysis on each sub-sample. From the set of results of the fits on sub-samples, we calculate the mean tension and potential strength, as well as their uncertainties as the standard deviations from the means (shown as error bars in Fig. 12). We present the outcome of this procedure for one patch with the low, and one patch with the high fluctuation amplitude (Figure 12a). We find the obtained uncertainties to be relatively small if the sub-samples are sufficiently long (about $\unit[25]{s})$, and the agreement between methods better at higher fluctuation amplitudes. The single measurement associated with the entire sequence typically falls within the uncertainty of the mean obtained with each method. This, together with the good reproducibility of the fit results between the sub-samples, strongly corroborates the reproducibility of results obtained by each method independently. Importantly, we find the uncertainties to be smaller than the uncertainty arising from the intrinsic experimental errors \cite{limozin2009,monzel2009, monzel2012}, and of the same magnitude as the uncertainty in determining the contour line in the phase diagram. Specifically, the small slope of the contour line suggests large uncertainties in the tension, while the large slope of the contour line is reflected in larger uncertainties in the potential strength. This is true even though determining the contour line is independent of the fitting procedures, at least in the case of spatial averaging and the time correlations. In the case of the latter,  the tension is most difficult to determine accurately, because the contour line is nearly flat.

Another instructive analysis is to compare the results obtained from different squares on the same vesicle, where at least the tension is expected to be same. This analysis is presented in Figure 12b for a vesicle that exhibits a statistically significant spread in mean fluctuation amplitudes of the patches, pointing to small variations in the substrate coating. The tension and the potential strength are found as the mean of values obtained from independent fits over eight squares (shown in the figure). Two short sub-samples (red and yellow symbols) are compared to one long sequence (black symbols). The results from each sub-sample reproduce the results over the whole sequence, supporting the finding discussed above (Fig 12a). Interestingly, if the average is performed only over patches with the similar mean fluctuation amplitude ($\unit[16\pm2]{nm^2}$), then the uncertainty in determining the effective potential with each method drops significantly (blue symbols in Fig 12b), suggesting that the substrate is similarly coated below these parts of the vesicle. Apart from further confirming the reproducibility of our approaches used for data analysis, this investigation is indicative of the uniformity of the substrate. Actually, one could infer that the sensitivity in determining the uniformity of the substrate coating obtained by measuring the membrane fluctuations is significantly larger than of other, more established methods. 

Finally, we analyze patches from different vesicles which were all prepared in the same way. Here one expects that patches with similar fluctuation amplitudes will yield similar values for the potential strength and tension, which is indeed the case (Fig. 12c). This agreement is very important as it clearly demonstrates the overall reproducibility of each approach independently, and justifies their individual application when suitable. In this context, the spatial averaging method is perhaps the most limited as it relies on relatively significant, apparent fluctuations of the membrane (weaker potentials and/or tensions). This is simply because at small fluctuation amplitudes the averaging curve (Fig. \ref{fig8}) flattens very quickly, which affects the sensitivity of the fit.

Out of all the three methods, obtaining the parameters from the shape may be technically most challenging, as it requires a non-trivial boundary problem to be solved numerically. As the first step in this procedure is determining the mean square fluctuation amplitude, the fitting result for $\sigma$ and $\gamma$ of this method are always exactly on the contour lines in Fig. \ref{fig11}. Despite its somewhat technical nature, this method points clearly to limitations of the commonly used harmonic approximation. Here, we showed that systematic values of the tension and the potential strength can be obtained only after making a more appropriate approximation for the direct membrane-substrate potential. Simple harmonic approximation would here systematically provide lower tensions and higher interaction potentials to provide a shape which reaches the minimum of the potential sufficiently fast. The difference in $\sigma$ and $\gamma$ may amount to a couple of orders of magnitude in a certain parameter range.

Even though all methods provided results within the same order of magnitude, the spatial averaging systematically provides the highest values of the potential strength and the smallest tension while the time correlation function provides the opposite, all with uncertainties that are smaller than the differences between the means associated with different methods. While the spatial averaging and the shape fitting rely exclusively on the equilibrium properties of the system (and provide similar results if the anharmonic potential is taken into account), the construction of the time correlation function requires the correct reconstruction of the hydrodynamic interactions of the membrane with the surrounding fluid, the latter based in $q$-dependent damping coefficients for the membrane close to the substrate \cite{seifert1994a}. The observed systematic deviations of about a factor of two suggest that, despite good agreement, a more in depth study of time correlations may be required before the behavior of the membrane can be fully resolved from the theoretical point of view. This analysis, which should combine modeling and experiments,  should clarify the role of a potential volume constraint, which was previously evoked in connection with the shape and fluctuations of adherent membranes [35]. On this note, our data suggest that changes in osmotic conditions will affect the volume below the patch, whereby we did not acquire any conclusive evidence that the volume constraint affects membrane fluctuations around an equilibrated shape. However, only a few modes are affected by the volume constraint in the square geometry, and hence different, more restrictive patterns should be used to fully understand its role.

\section{Conclusions}

The framework presented herein provides a set of tools for a systematic study of membrane-substrate interaction potentials, which is a key step toward the understanding of the decades-old puzzle, arising from inconsistencies in predictions and measurements of both the position of the minimum and the strength of the nonspecific potential. We have shown that this inconsistency can be removed to a large extent, if a more realistic potential is used to reconstruct the shape of the membrane. For this purpose, we have chosen the (4,2) Mie potential. Alternatively, we could have used the complete potential constructed by the superposition of the steric, hydration, van der Waals and other potentials. Such an approach would have the advantage of connecting the material properties of the system to the current description. While we have shown previously that it is possible to account for some of the qualitative behavior of the membrane within this superposition approach (change of the position of the minimum with modulating the membrane tension), we have also shown that the individual potentials are associated with a number of unknown parameters, including the Hamaker constant, which cannot be measured independently [56]. In contrast, the (4,2) Mie potential used here has the advantage of being defined by only two parameters, yet it captures the key features of the true effective potential, particularly around the minimum. Of course, very close and very far from the substrate, this potential is not correct. However, these two limits are irrelevant in practice because they are not visited by the membrane. Actually, the potential minimum is at relatively large distances from the substrate, and is associated with relatively small fluctuation amplitudes, which was also reported previously [27-29,48]. This may be a hint that the very approach of constructing the complete potential by superimposing the contributing potentials may be questionable, and that further studies of this potential are necessary. Our work here provides the key prerequisites for these next steps.

Irrespective of such details of the potential, we showed that the theoretical framework must be extended to account for anharmonic potentials. The first piece of evidence came from the reconstruction of the membrane shape. This method provided the membrane tension and potential strength consistent with the two methods relying on fluctuations only if the anharmonicity is taken into account. Some information about the functional form of the effective potential could be obtained by systematically inducing shape changes, yet the accuracy of such an approach is to be determined in the future.

The second piece of evidence for the anharmonic contributions came directly from measuring membrane fluctuations around the minimum (see Fig. 2). The latter can be reconstructed with great accuracy and the RICM is particularly well suited for these measurements. Again, systematic changes of system parameters would be necessary to gain deeper insight into the functional form of the effective potential, which will be a focus of further studies.

The true strength of our approach is, however, to insist on the consistency between various methods. Actually, it was exactly this requirement which pointed to the insufficiency of the harmonic description.The noteable discrepancy between equilibrium analysis and dynamic analysis suggests a further need for refinement of the theoretical treatment of hydrodynamic interactions.

In conclusion, determining the nonspecific potential between the membrane and another surface is a difficult problem, due to the coupling between the membrane tension, the steric repulsion and the direct interactions. Apart from putting into perspective the commonly used approximations, the work presented herein unambiguously showed that even small potentials affect the shape and the dynamics of the membrane significantly, suggesting that this potential needs to be treated earnestly in inter-membrane and membrane-substrate studies. One of the problems in the past has been the lack of consistency in experimental results. With this work, this predicament can be fully circumvented allowing us to now tackle the conceptual challenge of understanding this elusive, yet so effective potential.


\begin{acknowledgments}
We thank Susanne Fenz for useful discussions. ASS and TB acknowledge the funding of the Cluster of Excellence: Engineering of Advanced material at the University of Erlangen and the ERC StG 2013-337283 MembranesAct. CM is grateful for the support by the Deutsch-Französische Hochschule and the Exzellenzcluster Cellnetworks at the University of Heidelberg.
\end{acknowledgments}

\bibliography{referencesPRX.bib}

\end{document}